\documentclass[aps,showpacs,nofootinbib,superscriptaddress]{revtex4}
\usepackage{epsfig}
\usepackage{amsmath}
\newcommand{\ben}{\begin{eqnarray}}
\newcommand{\een}{\end{eqnarray}}
\newcommand{\nnu}{\nonumber\\}
\newcommand{\bef}{\begin{figure}[htb]\centering}
\newcommand{\eef}{\end{figure}}

\begin{document}
\title{Photon-tagged heavy meson production in high energy nuclear collisions}

\date{\today}

\author{Zhong-Bo Kang}
\email{zkang@bnl.gov}
\affiliation{RIKEN BNL Research Center,
                Brookhaven National Laboratory,
                Upton, NY 11973, USA}
                
\author{Ivan Vitev}
\email{ivitev@lanl.gov}
\affiliation{Los Alamos National Laboratory, 
                   Theoretical Division,
                   Los Alamos, NM 87545, USA}                

\begin{abstract}
We study the photon-triggered light and heavy meson production in both p+p and A+A collisions. 
We find that a parton energy loss approach that successfully describes inclusive hadron 
attenuation in nucleus-nucleus reactions at RHIC can simultaneously describe  well the 
experimentally determined photon-triggered light hadron fragmentation functions.
Using the same framework, we generalize our formalism to study photon-triggered 
heavy meson production. 
We find that the nuclear modification of photon-tagged  heavy meson fragmentation functions in 
A+A collision is very different from that of the photon-tagged light hadron case.  While photon-triggered 
light hadron fragmentation functions in A+A collisions are  suppressed relative to p+p, 
photon-triggered heavy meson fragmentation functions can be either enhanced 
or suppressed, depending on the specific kinematic region. The anticipated smaller energy 
loss for $b$-quarks manifests itself as a flatter photon-triggered $B$-meson fragmentation function 
compared to that for the $D$-meson case. We make detailed predictions for both RHIC and LHC energies. 
We conclude that a comprehensive comparative study of both photon-tagged light and 
heavy meson production can provide 
new insights in the details of the jet quenching mechanism. 
\end{abstract}


\maketitle

\section{Introduction}
High transverse momentum partons and their hadronic fragments are powerful and valuable 
probes of the 
high energy density matter created at the Relativistic Heavy Ion Collider (RHIC) and the Large Hadron 
Collider (LHC). These energetic partons are created in the early stage of the collisions and therefore 
provide  access to the space-time history of the transient hot and dense nuclear medium - the quark-gluon plasma (QGP) 
created in the heavy ion reactions. Specifically, as they propagate through the QGP, they interact 
with the medium and lose energy, a phenomenon known as 
``jet quenching''~\cite{Gyulassy:1993hr,Baier:1996sk,Zakharov:1997uu,Wiedemann:2000za,Gyulassy:2000er,Wang:2001ifa,Arnold:2002ja,Ovanesyan:2011xy}.

Experimentally, there has been tremendous progress in recent years~\cite{d'Enterria:2009am}
in establishing the jet quenching phenomenon from different observables, such as the suppression of 
single hadron  production~\cite{Adcox:2001jp,Adler:2002xw,Aamodt:2010jd}, 
dihadron correlations~\cite{Adler:2002tq,Adare:2010ry}, $\gamma$-hadron 
correlations~\cite{Adare:2009vd, Abelev:2009gu}, and ultimately the alteration of inclusive 
jet~\cite{Salur:2009vz} and dijet production~\cite{Aad:2010bu,Chatrchyan:2011sx}. Theoretically, many 
perturbative QCD-based models of jet quenching have been developed and used to extract the medium 
properties~\cite{Vitev:2009rd}, especially the QGP parton number and energy densities. They are able 
to successfully describe most of the experimental data and the measured nuclear modification 
factor $R_{AA}$ of single hadron production in particular. 

To study  jet quenching and parton energy loss in more detail and thus constrain the medium properties
better, more sophisticated observables have been proposed to improve our understanding of inelastic
parton interactions in the QGP. One such example 
is $\gamma$-triggered hadron production and correlations~\cite{Wang:1996yh}: one studies jet quenching by 
measuring the $p_T$ distribution of charged hadrons in the opposite direction of a trigger direct photon. 
Since the photon does not interact with the dense medium,
its energy approximately reflects that of the initial away-side parton before energy loss up to corrections 
at next-to-leading order in $\alpha_s$. One can therefore study the effective medium modification of 
the jet fragmentation function. Earlier studies on the $\gamma$-triggered light hadron 
correlations~\cite{Zhang:2009rn,Qin:2009bk,Renk:2009ur} seem to be roughly compatible with the 
experimental data.

Although jet quenching has been successful in describing most of the experimental findings, there are still 
remaining puzzles. Resolving these puzzles will eventually uncover the exact underlying mechanism for the suppression
of leading particles and jets. One well-known difficulty is related to the fact that the $c$ and $b$-quark parton-level 
energy loss in the QGP has not been sufficient in the past to explain the large suppression of 
non-photonic $e^{+}+e^{-}$ 
measured at RHIC~\cite{Dokshitzer:2001zm,Wicks:2005gt}. These non-photonic electrons come from the semileptonic 
decays of $D$ and $B$ mesons. In the framework of perturbative QCD, collisional dissociation of heavy-mesons 
in the QGP has been suggested as an additional suppression mechanism~\cite{Adil:2006ra,Sharma:2009hn}. To  
distinguish with confidence between theoretical models one hopes to have a direct measurement of the heavy meson
cross sections in both p+p and A+A collisions. Such direct measurements are finally becoming  available from 
both RHIC and the LHC experiments.

The main goal of this paper is to investigate new experimental observables, such as $\gamma$-triggered 
hadron production, and make simultaneous predictions for both light and heavy meson final states.
Naturally, we will first focus on the standard parton energy loss mechanism in the QGP. More specifically,
we present detailed studies of both $\gamma$-triggered away-side light hadron and heavy meson spectra 
in heavy ion collisions. Within the same energy loss formalism - the Gyulassy-Levai-Vitev (GLV) approach - 
we study how the QGP medium affects the $\gamma$-triggered light and heavy meson effective fragmentation 
functions. We find that the nuclear modification factor $I_{AA}$ behaves very differently depending on the
parent parton mass: $I_{AA}$ is considerably suppressed for light hadrons ($I_{AA} < 1$),  whereas it can be 
suppressed ($I_{AA} < 1$)  or enhanced ($I_{AA} > 1$) for heavy mesons,  depending on the kinematic region, due to the different 
shape of the heavy quark fragmentation functions. Furthermore, we find that the nuclear modification $I_{AA}$ 
is flatter in the $B$-meson case compared to that of $D$-mesons due to the smaller energy loss for $b$-quarks 
in the non-asymptotic (finite energy) case.  Thus, a comparative study of $\gamma$-triggered light 
and heavy meson correlations can be a very useful probe of the physics that underlays the 
experimentally established jet quenching.

The rest of our paper is organized as follows: in Sec.~II we present the relevant formalism for both 
photon-tagged light hadron and heavy meson production. In Sec.~III we present our phenomenological studies. 
We first compare our calculation to the experimental data on photon+light hadron correlations. Then, we 
make predictions for photon-triggered light and heavy meson production relevant to both RHIC and LHC experiments 
and discuss the differences in the observed nuclear modification. We conclude our paper in Sec.~IV.

\section{Photon-tagged light and heavy meson production}
In this section we present the relevant formalism for the photon-triggered light and heavy meson production in 
nucleon-nucleon collisions. These formula will then be used in the phenomenological studies in the next section.

\subsection{Photon-tagged light hadron production}
Within the framework of the collinear perturbative QCD factorization approach, the lowest order (LO) differential 
cross section for back-to-back photon-light-hadron production can be obtained from the partonic processes: 
$q\bar{q}\to \gamma g$ and $qg\to \gamma q$ and is given by~\cite{Owens:1986mp}
\ben
\frac{d\sigma^{\gamma h}_{NN}}{dy_\gamma dy_h dp_{T_\gamma}dp_{T_h}}=
\frac{2\pi\alpha_{em}\alpha_s}{S^2}
\sum_{a,b,c}
D_{h/c}(z_T) \frac{f_{a/N}(x_a)f_{b/N}(x_b)}{x_a x_b}\overline{|M|}^2_{ab\to \gamma c}\, ,
\label{light}
\een
where $S$ is the squared center of mass energy of the hadronic collisions, and $z_T$, $x_a$, and $x_b$ are given by
\ben
z_T=\frac{p_{T_h}}{p_{T_\gamma}},
\qquad
x_a=\frac{p_{T_\gamma}}{\sqrt{S}}\left(e^{y_\gamma}+e^{y_h}\right),
\qquad
x_b=\frac{p_{T_\gamma}}{\sqrt{S}}\left(e^{-y_\gamma}+e^{-y_h}\right).
\een
Here, $y_\gamma$ and $p_{T_\gamma}$ ($y_h$ and $p_{T_h}$) are the rapidity and transverse momentum of the photon 
(away-side hadron), respectively. We denote by $f_{a,b/N}(x_{a,b})$ the distribution functions of 
partons $a,b$ in the nucleon and  $D_{h/c}(z)$ is the fragmentation function of parton 
$c$ into hadron $h$. $\overline{|M|}^2_{ab\to \gamma c}$ are the squared 
matrix elements for $ab\to \gamma c$ partonic processes, and are given by
\ben
\overline{|M|}^2_{qg\to \gamma q}&=&e_q^2\frac{1}{N_c}\left[-\frac{s}{t}-\frac{t}{s}\right],
\qquad
\overline{|M|}^2_{gq\to \gamma q}=e_q^2\frac{1}{N_c}\left[-\frac{s}{u}-\frac{u}{s}\right],
\\
\overline{|M|}^2_{q\bar q\to \gamma g}&=&\overline{|M|}^2_{\bar q q\to \gamma g}
=e_q^2\frac{N_c^2-1}{N_c^2}\left[\frac{t}{u}+\frac{u}{t}\right],
\een
where $s, t, u$ are the partonic Mandelstam variables, $e_q$ is the fractional electric charge of the light quark, and $N_c=3$ is the number of colors.

Experimentally, one typically defines the so-called $\gamma$-triggered fragmentation function
\ben
D_{NN}^{\gamma h}(z_T)&=&\frac{\int dy_\gamma dy_h dp_{T_\gamma} p_{T_\gamma}\frac{d\sigma^{\gamma h}_{NN}}{dy_\gamma 
dy_h dp_{T_\gamma}dp_{T_h}}}{\int dy_\gamma dp_{T_\gamma} \frac{d\sigma_{NN}^{\gamma}}{dy_\gamma dp_{T_\gamma}}}
\label{pplight}
\een
for nucleon-nucleon collisions and a similar $D_{AA}^{\gamma h}(z_T)$ per binary scattering for 
A+A collisions. The denominator defines the normalization and is given by
\ben
\frac{d\sigma_{NN}^{\gamma}}{dy_\gamma dp_{T_\gamma}}=\sum_h \int dy_h dp_{T_h}
\frac{d\sigma^{\gamma h}_{NN}}{dy_\gamma dy_h dp_{T_\gamma}dp_{T_h}} , 
\een
which is just the cross section for direct photon production. To quantify the modification of $\gamma$-triggered 
fragmentation function in A+A collisions relative to that in nucleon-nucleon collisions due to the jet quenching, 
one introduces the nuclear modification factor, 
\ben
I_{AA}^{\gamma h}(z_T)=\frac{D_{AA}^{\gamma h}(z_T)}{D_{NN}^{\gamma h}(z_T)}.
\een
Note that, in spite of their suggestive name, $\gamma$-triggered fragmentation functions are derived from the observed 
away-side meson distribution and thus reflect all input in a pQCD calculation - the parton distributions, the hard
scattering cross sections, the medium-induced radiative correction (or parton energy loss) and the parton decay 
probabilities - not only the fragmentation process  itself.

\subsection{Photon-tagged heavy meson production}
Within the perturbative QCD factorization approach there have been different ways to calculate heavy quark production. 
One of them is called a fixed-flavor-number scheme (FFNS)~\cite{Frixione:1994dv,Frixione:1995qc,Stratmann:1994bz} and is based 
on the assumption that the gluon and the three light quarks ($u, d, s$) are the only active partons. The heavy quark appears 
only in the final state and is produced in the hard scattering process of light partons.  The heavy quark mass $m$ is explicitly 
retained throughout the calculation. The other approach is called a variable-flavor-number scheme 
(VFNS)~\cite{Berger:1995qe,Bailey:1996px,Stavreva:2009vi} and describes the heavy quark as a massless parton of
density $f_{Q/N}(x, \mu^2)$ in the nucleon, with the boundary condition $f_{Q/N}(x, \mu^2)=0$ for $\mu\leq m$.  Thus, 
the heavy quark mass $m$ is set to zero in the short-distant partonic cross 
section\footnote{Strictly speaking, this is the so-called zero-mass variable-flavor-number scheme (ZM-VFNS). There is 
also general-mass variable-flavor-number scheme (GM-VFNS) which combines the virtues of the FFNS and the ZM-VFNS, 
though more complicated.}.

In this paper, we are particularly interested in studying the jet quenching effects for both light and heavy mesons. 
The different amount of energy loss during quark propagation in the medium created in heavy ion collisions is due to 
the mass difference between light and heavy quarks, at least in perturbative QCD. 
With this in mind, we want to keep explicitly the heavy quark mass in our calculation. In other words, we favor the 
FFNS scheme for the project at hand. 
Within this scheme, the photon-tagged heavy mesons are produced through the following partonic processes 
- (a) quark-antiquark annihilation: $q\bar{q}\to Q\bar{Q}\gamma$; (b) gluon-gluon fusion: $gg\to Q\bar Q \gamma$. 
The sample Feynman diagrams are given in Fig.~\ref{feynfig}. It is important to realize that in the FFNS scheme 
photon+heavy quark events are generated to LO by the $2\to 3$ processes, to be compared to the usual 
$2\to 2$ processes for the photon+light hadron events at LO.
\bef
\psfig{file=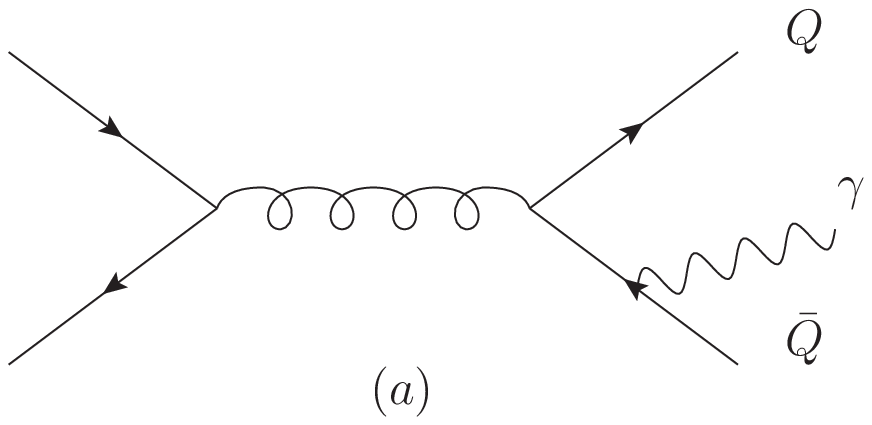,width=1.8in}
\hskip 0.3in
\psfig{file=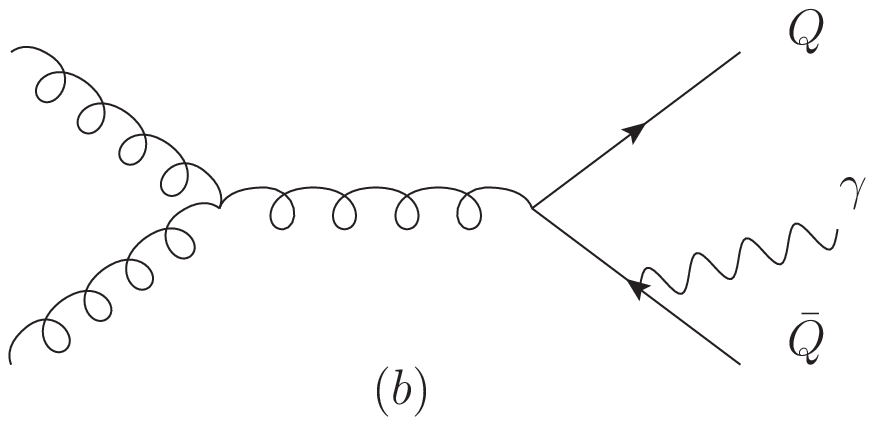,width=1.8in}
\hskip 0.3in
\psfig{file=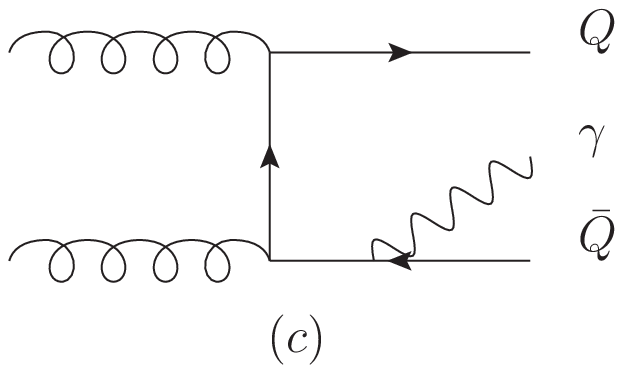,width=1.5in}
\caption{Sample Feynman diagrams at leading order for direct photon plus heavy quark production: (a) for light quark-antiquark annihilation $q\bar q\to Q\bar Q \gamma$, (b) and (c) for gluon-gluon fusion $gg\to Q\bar Q \gamma$.}
\label{feynfig}
\eef

The differential cross section for photon-tagged heavy meson production takes the following form~\cite{Stratmann:1994bz}:
\ben
\frac{d\sigma^{\gamma H}_{NN}}{dy_\gamma dy_h dp_{T_\gamma}dp_{T_H}}
&=&\frac{\alpha_{em}\alpha_s^2}{2\pi S}\sum_{ab}\int \frac{dz}{z^2}D_{H/Q}(z)
\int \frac{dx_a}{x_a} f_{a/N}(x_a) \int d\phi \frac{1}{x_b} f_{b/N}(x_b)
\nnu
&&
\times
\frac{p_{T_\gamma}p_{T_H}}{x_aS-\sqrt{S}(p_{T_\gamma} e^{y_\gamma}+m_{T_Q}e^{y_H})}
\overline{|M|}_{ab\to Q\bar Q\gamma},
\label{heavy}
\een
where $D_{H/Q}(z)$ is the heavy quark $Q$ to heavy meson $H$ fragmentation function and 
$\phi=\phi_h-\phi_\gamma \in[\pi/2, 3\pi/2]$ is the azimuthal angle between the triggered photon and the associated 
heavy meson on the away side. We denote by $m_{T_Q}=\sqrt{p_{T_Q}^2+m^2}$  the transverse mass and by $p_{T_Q}=p_{T_H}/z$ 
the transverse momentum for the heavy quark. The parton momentum fraction $x_b$ is fixed by the kinematics and 
has the following form
\ben
x_b=\frac{x_a\sqrt{S}(p_{T_\gamma}e^{-y_\gamma}+m_{T_Q}e^{-y_H})
+2p_{T_\gamma}p_{T_Q}\cos\phi-p_{T_\gamma}m_{T_Q}(e^{y_\gamma-y_H}+e^{y_H-y_\gamma})}
{x_a S-\sqrt{S}(p_{T_\gamma}e^{y_\gamma}+m_{T_Q}e^{y_H})}.
\een

The partonic matrix elements $\overline{|M|}_{ab\to Q\bar Q\gamma}$ can be calculated in perturbative QCD. 
Note that the cross section does not produce any singularity since the heavy quarks have been taken to be massive 
explicitly. Therefore no regularization is needed despite the fact that we are actually dealing with 
a $2\to 3$ process. The cross section for photoproduction of heavy quarks has been published in~\cite{Ellis:1987qy}. 
By using crossing symmetry, we can infer the spin-averaged matrix elements in our case. We list the results 
here for completeness. For the quark-antiquark annihilation channel 
$q(p_1)+\bar q(p_2)\to Q(p_3)+\bar Q(p_4)+\gamma(p_5)$ we have
\ben
\overline{|M|}^2_{q\bar q\to Q\bar Q\gamma}=\frac{N_c^2-1}{N_c^2}\left[e_Q^2A_1+e_q^2 A_2 + e_Q e_q A_3\right],
\een
where $e_Q$ is the fractional electric charge of the heavy quark, and  
the coefficients $A_1$, $A_2$, and $A_3$ are given by
\ben
A_1&=&\frac{\left(p_{24}^2+p_{23}^2+p_{14}^2+p_{13}^2+m^2 s_{34}\right)p_{34}}
{p_{45} p_{35} p_{12}  s_{34}}
+\frac{m^2 \left(p_{25}^2+p_{15}^2\right)}{2 p_{12}^2 p_{45}p_{35}}
\nnu
&&
 +\frac{m^2\left(p_{24}^2+p_{14}^2+p_{23}^2+p_{13}^2-p_{12}s_{34}\right)}
 {p_{12}^2 s_{34}}\left[\frac{1}{p_{45}}+\frac{1}{p_{35}}\right]
\nnu
&&
-\frac{m^2}{2p_{12}^2} \left[\frac{p_{24}^2+p_{14}^2+m^2p_{12}}{p_{35}^2}
+\frac{p_{23}^2+p_{13}^2+m^2p_{12}}{p_{45}^2}
\right],
\\
A_2&=&\frac{p_{24}^2+p_{23}^2+p_{14}^2+p_{13}^2+2 m^2p_{12}}
{p_{25} p_{15}  s_{34}}
+\frac{2 m^2\left(p_{25}^2+p_{15}^2\right)}{p_{25} p_{15} s_{34}^2},
\\
A_3&=&\frac{p_{24}^2+p_{23}^2+p_{14}^2+p_{13}^2+m^2 (p_{12}+\frac{1}{2}s_{34})}
{p_{12}  s_{34}}
\left[\frac{p_{24}}{p_{45}p_{25}}+\frac{p_{13}}{p_{35}p_{15}}-\frac{p_{14}}{p_{45}p_{15}}
-\frac{p_{23}}{p_{35}p_{25}}\right]
\nnu
&&
+\frac{2m^2}{p_{12}s_{34}}\left[\frac{p_{13}-p_{23}}{p_{45}}+\frac{p_{24}-p_{14}}{p_{35}}\right].
\een
Here, we have used the notation $p_{ij}=p_i\cdot p_j$, $s_{34}=(p_3+p_4)^2$.
For the gluon-gluon fusion channel $g(p_1)+g(p_2)\to Q(p_3)+\bar Q(p_4)+\gamma(p_5)$  we have
\ben
\overline{|M|}^2_{g g\to Q\bar Q\gamma}=-\frac{1}{N_c(N_c^2-1)}e_Q^2
\left[\left(R_{QED}+\mbox{11 perm's}\right)+N_c^2\left(R_{KF}+\mbox{3 perm's}\right)\right].
\een
The $R_{QED}$ has to be summed over 12 permutations corresponding to the 6 permutations of the momenta 
$p_1$, $p_2$, $-p_5$ and two permutations of the momenta $p_1$ and $p_2$. $R_{KF}$ has to be summed over 
4 permutations corresponding to the interchange of $p_1$, $p_2$ and $p_3$, $p_4$. One such 
expression for $R_{QED}$ is given by
\ben
R_{QED}&=&\frac{s_{34} p_{24}^2}{8p_{14}p_{13}p_{45}p_{35}}
\nnu
&&
-\frac{m^2}{2p_{35}}\left[\frac{1}{p_{14}}\left(\frac{s_{34}}{2p_{13}}-\frac{p_{23}}{p_{14}}+3\right)
+\frac{p_{45}-p_{23}+p_{14}}{2p_{24}p_{13}}-\frac{1}{p_{45}}-\frac{3}{2p_{13}}
\right]
\nnu
&&
-\frac{m^4}{2p_{35}p_{14}}\left[
\frac{1}{2p_{13}}\left(\frac{p_{13}-3p_{14}+2p_{45}}{p_{24}}+\frac{3p_{24}}{p_{45}}-4\right)
+\frac{2}{p_{35}}
\right]
\nnu
&&
+\frac{m^6}{p_{35}p_{14}}\left[\frac{1}{2p_{13}}\left(\frac{1}{2p_{45}}-\frac{1}{p_{24}}\right)
+\frac{1}{2p_{14}}\left(\frac{1}{2p_{35}}-\frac{1}{p_{23}}\right)
\right],
\een
and $R_{KF}$ is given by 
\ben
R_{KF}&=&-\frac{1}{2p_{12}}\left[\frac{p_{45}^2}{p_{13}p_{24}}
+\frac{p_{14}^2}{p_{45}p_{35}}\left(\frac{p_{14}}{p_{24}}+\frac{p_{13}}{p_{23}}\right)
\right]
\nnu
&&
+\frac{m^2}{2p_{12}^2}\left[1-\frac{2p_{14}}{p_{35}}\left(\frac{p_{14}}{p_{35}}+\frac{p_{23}}{p_{45}}\right)\right]
+\frac{m^2}{4p_{14}p_{23}}\left[-\frac{2p_{45}}{p_{23}}+\frac{2(p_{45}+p_{35})}{p_{12}}-7\right]
\nnu
&&
+\frac{m^2}{2p_{35}p_{14}}\left[\frac{p_{45}-p_{12}}{p_{23}}+\frac{p_{23}-p_{12}}{p_{45}}
-\frac{p_{23}}{p_{14}}+\frac{p_{12}}{p_{35}}+7\right]
\nnu
&&
+\frac{m^2}{p_{35}}\left[\frac{1}{p_{12}}\left(\frac{p_{14}-p_{45}}{p_{23}}+\frac{2p_{14}}{p_{35}}
+\frac{2p_{14}}{p_{45}}-2\right)-\frac{1}{p_{45}}-\frac{3}{2p_{35}}
\right]
\nnu
&&
+\frac{m^4}{p_{35}p_{23}}\left[\frac{1}{p_{12}}\left(\frac{p_{45}}{p_{14}}+\frac{p_{14}}{p_{45}}\right)
+\frac{1}{2p_{45}}\left(\frac{p_{12}}{p_{14}}+2\right)\right]
\nnu
&&
+\frac{m^4}{p_{35}}\left[\frac{1}{p_{12}}\left(-\frac{1}{p_{23}}-\frac{1}{p_{35}}+\frac{1}{p_{14}}-\frac{2}{p_{45}}\right)-\frac{1}{p_{14}}\left(\frac{1}{p_{14}}-\frac{1}{p_{35}}+\frac{2}{p_{23}}\right)\right]
\nnu
&&
+\frac{m^4}{p_{14}p_{23}}\left(\frac{1}{2p_{12}}+\frac{1}{p_{14}}\right)
-\frac{m^6}{p_{14}^2}\left[\left(\frac{1}{2p_{35}}+\frac{p_{14}}{2p_{23}p_{45}}\right)^2
+\frac{1}{p_{23}}\left(\frac{1}{4p_{23}}-\frac{1}{p_{35}}\right)
\right].
\een
With the cross sections at hand, the experimentally accessible $\gamma$-triggered heavy meson 
fragmentation function is defined as
\ben
D_{NN}^{\gamma H}(z_T)&=&\frac{\int dy_\gamma dy_H dp_{T_\gamma} p_{T_\gamma}
\frac{d\sigma^{\gamma H}_{NN}}{dy_\gamma dy_H dp_{T_\gamma}dp_{T_H}}}{\int dy_\gamma dp_{T_\gamma} 
\frac{d\sigma_{NN}^{\gamma Q}}{dy_\gamma dp_{T_\gamma}}},
\een
where the denominator is the differential cross section of photon and heavy meson production summed over all the heavy mesons
\ben
\frac{d\sigma_{NN}^{\gamma Q}}{dy_\gamma dp_{T_\gamma}}
=\sum_H \int dy_H dp_{T_H} 
\frac{d\sigma^{\gamma H}_{NN}}{dy_\gamma dy_H dp_{T_\gamma}dp_{T_H}}.
\een
It can be written as
\ben
\frac{d\sigma_{NN}^{\gamma Q}}{dy_\gamma dp_{T_\gamma}}
&=&
\frac{\alpha_{em}\alpha_s^2}{2\pi S}\sum_{ab}\int dy_Q dp_{T_Q}\int \frac{dx_a}{x_a} f_{a/N}(x_a) \int d\phi \frac{1}{x_b} f_{b/N}(x_b)
\nnu
&&
\times
\frac{p_{T_\gamma}p_{T_Q}}{x_aS-\sqrt{S}(p_{T_\gamma} e^{y_\gamma}+m_{T_Q}e^{y_Q})}
\overline{|M|}_{ab\to Q\bar Q\gamma},
\een
which is just part of the inclusive photon production cross section that has  $\gamma+Q$ events.

Likewise, we can  define a  $\gamma$-triggered heavy meson fragmentation function $D_{AA}^{\gamma H}(z_T)$ in A+A collisions  
and the nuclear modification factor $I_{AA}^{\gamma H}(z_T)=D_{AA}^{\gamma H}(z_T)/D_{NN}^{\gamma H}(z_T)$. In the next section 
we will study how these fragmentation functions behave in both p+p and A+A collisions and, most importantly, 
how they are modified in A+A collisions due to  jet quenching.

\section{Phenomenological studies in p+p and A+A collisions}
In this section we will study the $\gamma$-triggered light hadron and heavy meson production in 
both p+p and A+A collisions. 
Final-state medium-induced radiative corrections are process dependent -  they depend on the parameters of 
the QGP medium and, consequently, on the details of the heavy ion collision. However, in QCD they factorize from 
the hard scattering cross section~\cite{Ovanesyan:2011xy} and enter the physical observables as a standard 
integral  convolution. From the point of view of phenomenological applications, it is  convenient to change the 
order of integration and include parton energy loss as an 
effective replacement of the fragmentation function $D_{h/c}(z)$~\cite{Vitev:2002pf}:
\ben
D_{h/c}(z) \Rightarrow \int_0^{1-z} d\epsilon\, P(\epsilon) \,
\frac{1}{1-\epsilon} D_{h/c}\left(\frac{z}{1-\epsilon}\right).
\label{quenching}
\een
Here, $P(\epsilon)$ is the probability distribution for a parton to lose a fraction of its energy 
$\epsilon=\sum_i\frac{\Delta E_i}{E}$ due to multiple gluon emission. In our calculation 
$P(\epsilon)$ is obtained using the GLV formalism for evaluating the medium-induced gluon 
bremsstrahlung~\cite{Gyulassy:2000fs,Vitev:2007ve} and the Poisson approximation. 
Note that it is not mandatory to employ Eq.~(\ref{quenching}) in jet quenching phenomenology. One can
alternatively evaluate the attenuated  partonic cross section prior to fragmentation and obtain
identical results~\cite{Sharma:2009hn}.

We take into account parton energy loss in $\gamma$-hadron correlation in high energy heavy-ion collisions
as in  Eq.~(\ref{quenching}) for simplicity. We do not perform a new evaluation of the energy loss since we aim at 
a direct and consistent comparison between different final states in the same jet quenching formalism. 
Instead, we employ the differential medium-induced bremsstrahlung distributions and energy loss probabilities
$P(\epsilon)$ for light partons and heavy quarks obtained in~\cite{Sharma:2009hn,Neufeld:2010fj}. An illustration 
of these probabilities 
\ben
\int_0^1 P(\epsilon) d\epsilon =1, \qquad \int_0^1 \epsilon \, P(\epsilon) d\epsilon 
=  \left\langle \frac{\Delta E}{E}  \right\rangle , 
\qquad \epsilon=\sum_i\frac{\Delta E_i}{E}
\label{prob}
\een
is given in Fig.~\ref{peps}. Note that, in the probabilistic application of parton energy loss, 
larger $\langle \Delta E / E \rangle$ corresponds to a distribution skewed toward large $\epsilon$
and smaller  $\langle \Delta E / E \rangle$ corresponds to a distribution skewed toward small $\epsilon$.
All results include an average over the jet production points distributed according to the binary
collision density in an optical Glauber model. These energetic jets propagate through the Bjorken 
expanding medium that follows the number of participants density~\cite{Sharma:2009hn,Neufeld:2010fj}.  
The left panel of Fig.~\ref{peps} corresponds to central Au+Au collisions at RHIC and parton energy 
$E = 10$~GeV and the right panel of Fig.~\ref{peps} corresponds to central Pb+Pb collisions at 
the LHC and parton energy  $E=25$~GeV. 

The difference between light quarks and gluons arises from the quadratic Casimir, or average squared
color charge, of the parton. Note that the naive relation $\langle \Delta E  \rangle_g  = C_A / C_F
\langle \Delta E  \rangle_q $ is approximately fulfilled only if $\langle \Delta E  \rangle_{q,g}  \ll E$.
In the realistic case,  $\langle \Delta E  \rangle_g  < C_A / C_F
\langle \Delta E  \rangle_q $ and the deviation form the naive scaling can be significant.
On the other hand, the difference between light and heavy quarks is related to the quark 
mass~\cite{Dokshitzer:2001zm,Wicks:2005gt,Vitev:2007ve}. In the very high energy limit, for coherent 
medium-induced  bremsstrahlung, the mass dependence of parton energy loss disappears. The calculations 
that we present in this paper are not in this asymptotic limit. Furthermore, for partons of effective mass 
$ gT \sim m_c$, where $g$ is the coupling constant and $T$ is the temperature of the quark-gluon plasma
created in heavy ion collisions, there is very little difference between the light quarks and the charm
quarks. This is clearly seen in both panels of Fig.~\ref{peps} and well understood. For the much heavier
bottom quarks $m_b \gg m_c \sim gT $, however, the energy loss is noticeably smaller and $P(\epsilon)$
is skewed toward smaller values of $\epsilon$.

\bef
\psfig{file=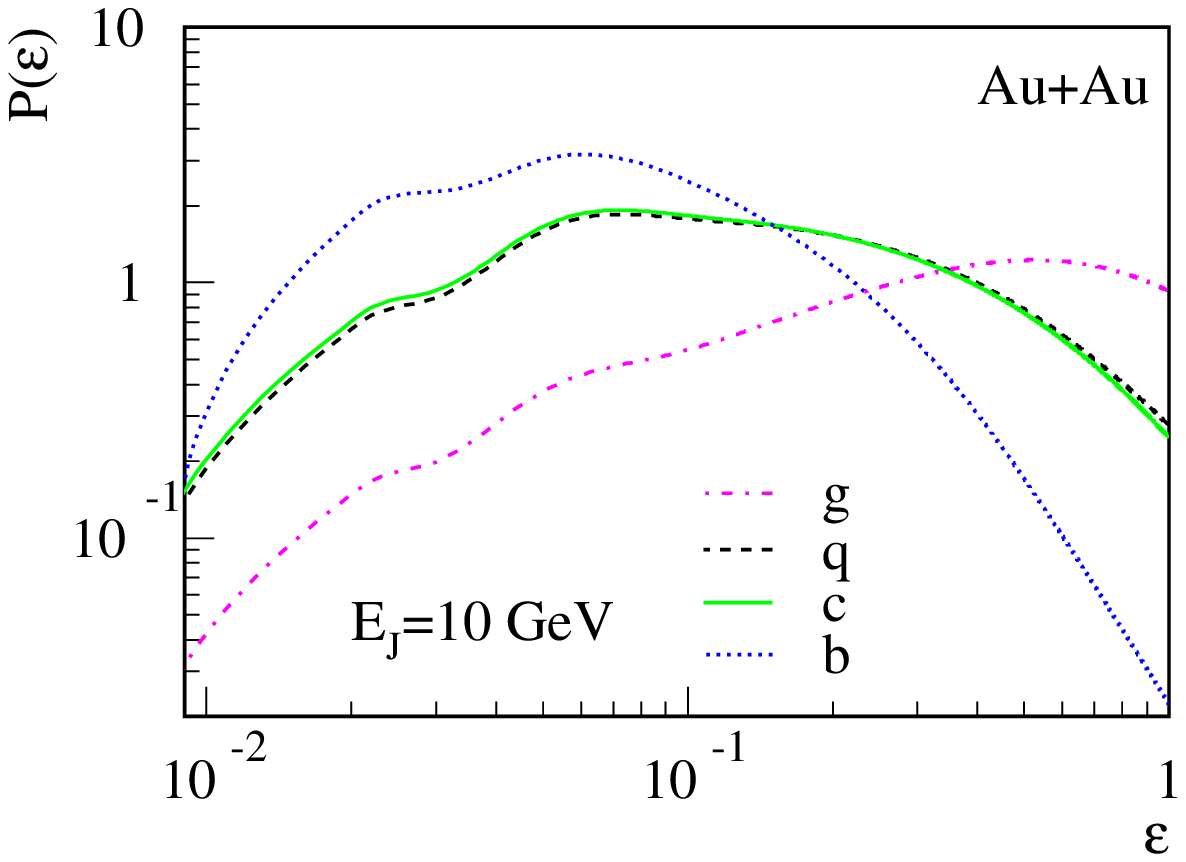,width=2.7in}
\hskip 0.4in
\psfig{file=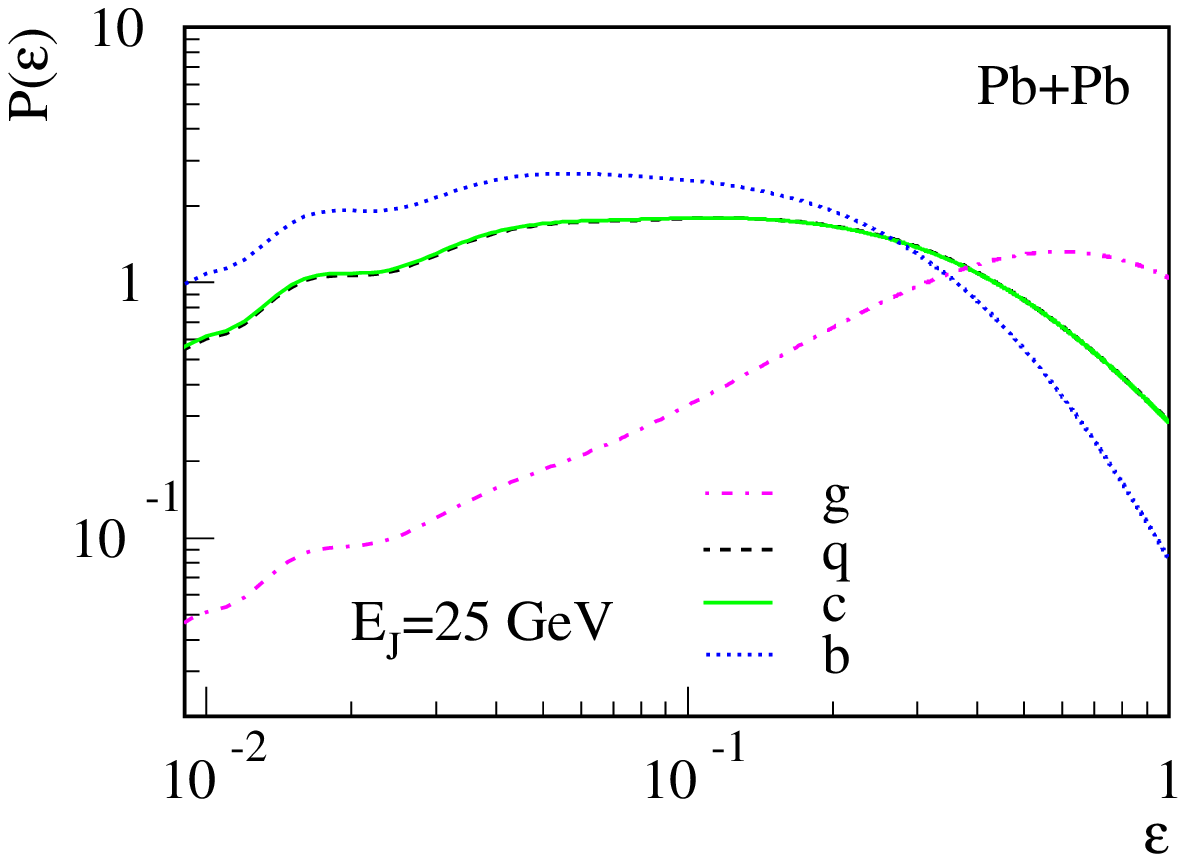,width=2.7in}
\caption{The probability distribution $P(\epsilon)$ for partons to lose a fraction $\epsilon$  of their energy
is shown for gluons (dot-dashed), light quarks (dashed), charm quarks (solid) and bottom quarks (dotted). 
Left  panel: central Au+Au collisions at RHIC and parton energy $E_J = 10$~GeV. 
Right  panel: central Pb+Pb collisions at the LHC and parton energy $E_J = 25$~GeV. }
\label{peps}
\eef

\subsection{Comparison to  experimental data: $\gamma$-triggered light hadron production}
In this subsection we will compare our calculations for $\gamma$-triggered light hadron production to the experimental 
data at RHIC. We will use CTEQ6 parton distribution function~\cite{Pumplin:2002vw} and deFlorian-Sassot-Stratmann
(DSS) light hadron fragmentation function~\cite{deFlorian:2007aj}.

In Fig.~\ref{starphenix} (left and top right panels) we show our results compared to both PHENIX~\cite{Adare:2009vd} 
and STAR data~\cite{Abelev:2009gu}. For p+p collisions we use the definition given in Eq.~(\ref{pplight}). As seen 
from these figures, the perturbative QCD calculation gives a very good descriptions of the experimental data. 
For Au+Au collisions we use the same energy loss distributions that were used to describe  single hadron suppression. 
As can be seen from Fig.~\ref{starphenix}, the calculated $D_{AA}^{\gamma h}(z_T)$ are in  good agreement with the 
experimental data with the possible exception of the large $z_T \rightarrow 1$ region in the STAR data. This part
of phase space is also inherently difficult to measure accurately. Such agreement gives an independent confirmation 
of the jet quenching mechanism in light hadron production.

\bef
\psfig{file=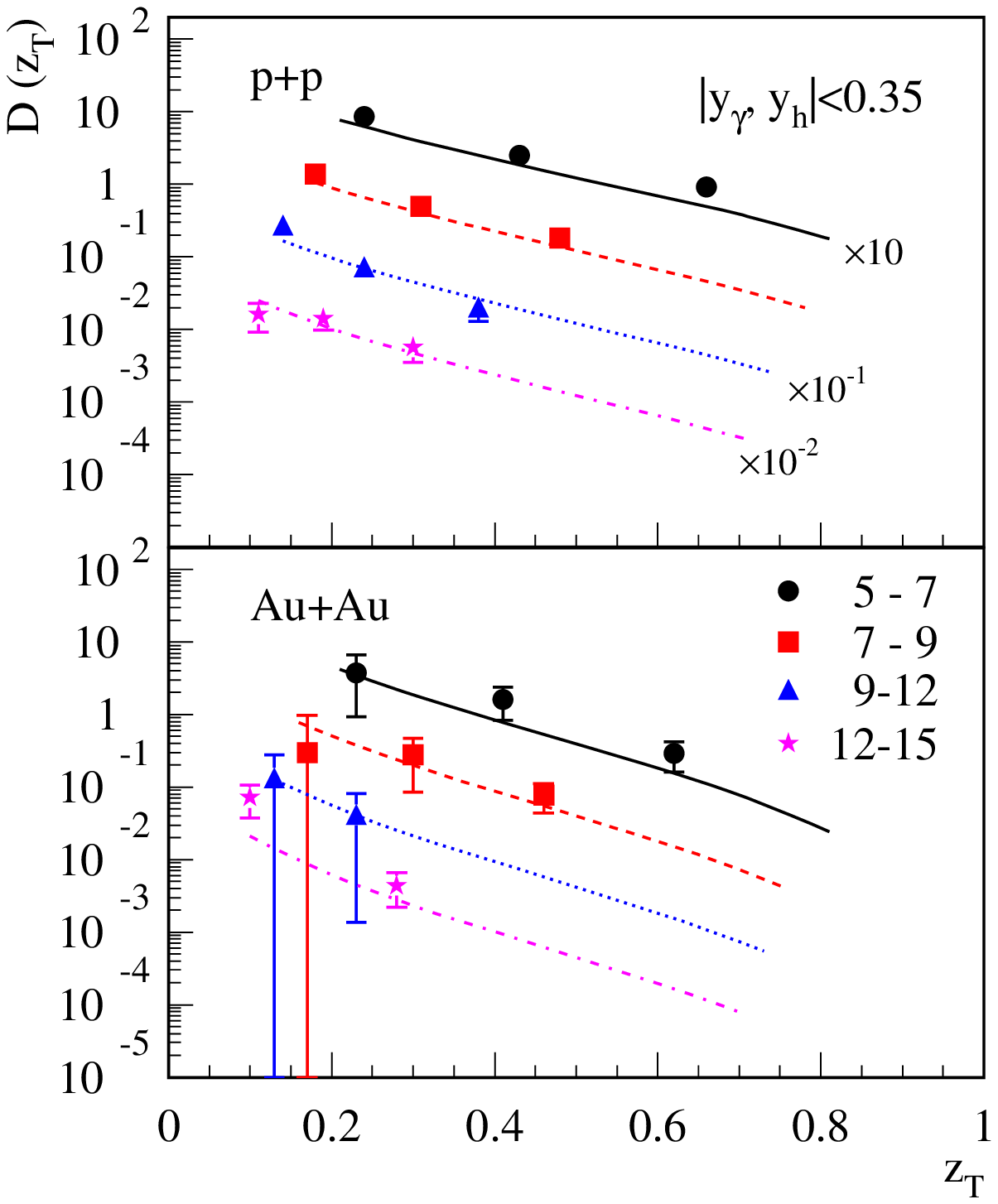, width=2.8in}
\hskip 0.4in
\psfig{file=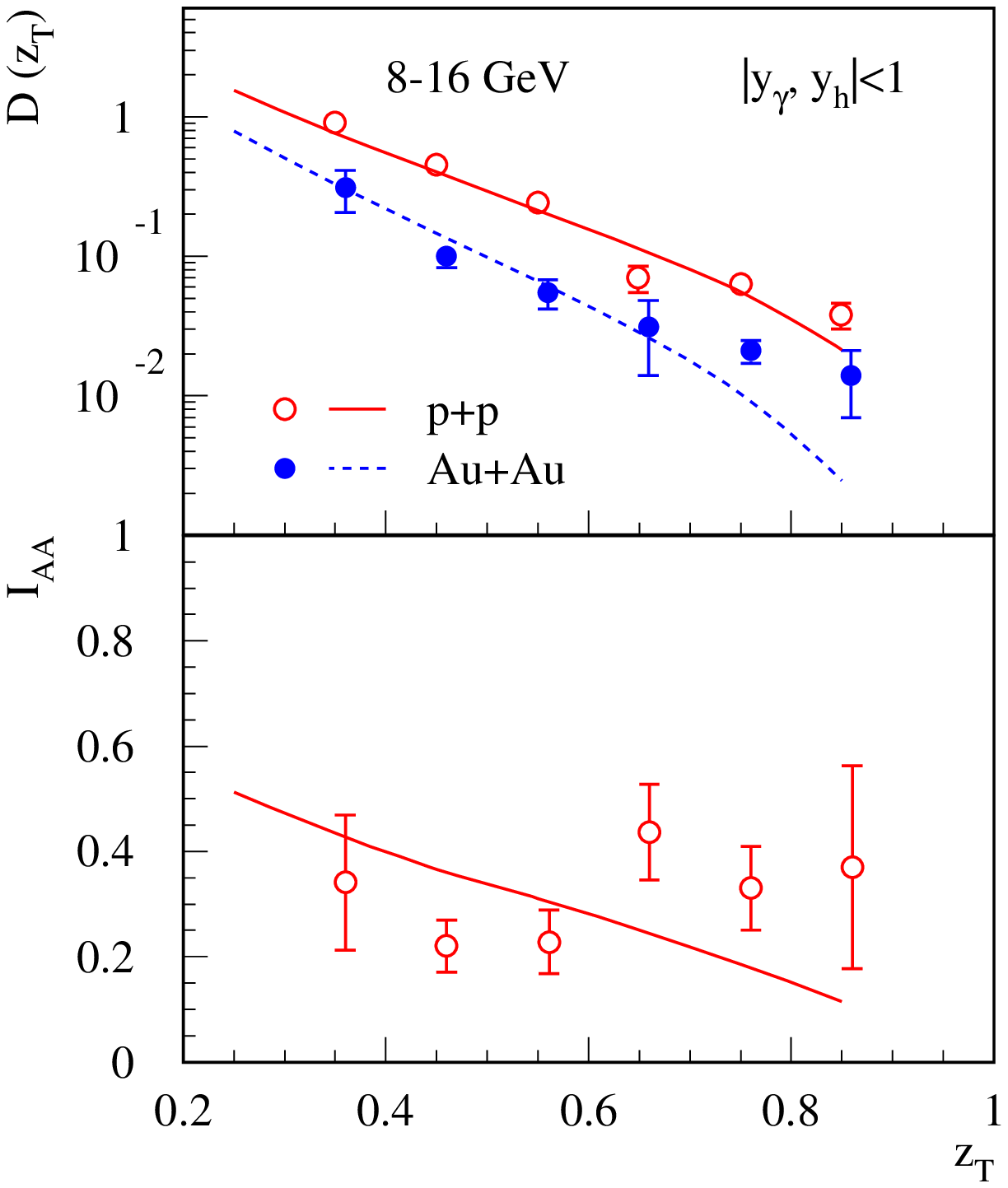, width=2.9in}
\caption{$\gamma$-triggered fragmentation functions $D(z_T)$ (left) and (top right) are plotted as a function of $z_T$ 
at $\sqrt{S_{NN}}=200$ GeV. Left: top for p+p collisions, bottom for Au+Au collisions. Photon and hadron 
rapidities are integrated over $[-0.35, 0.35]$, while the trigger particle (the photon) momentum has been integrated 
over $[5, 7]$, $[7,9]$, $[9, 12]$, and $[12, 15]$ GeV from top to bottom. Data is from PHENIX~\cite{Adare:2009vd}. 
Right: top for $D(z_T)$ - solid for p+p collision, and dashed for Au+Au collision; bottom for nuclear modification factor 
$I_{AA}$. The photon and hadron rapidities are integrated over $[-1,1]$, and the photon momentum is integrated over 
$[8, 16]$ GeV. Data is from STAR~\cite{Abelev:2009gu}.}
\label{starphenix}
\eef
\bef
\psfig{file=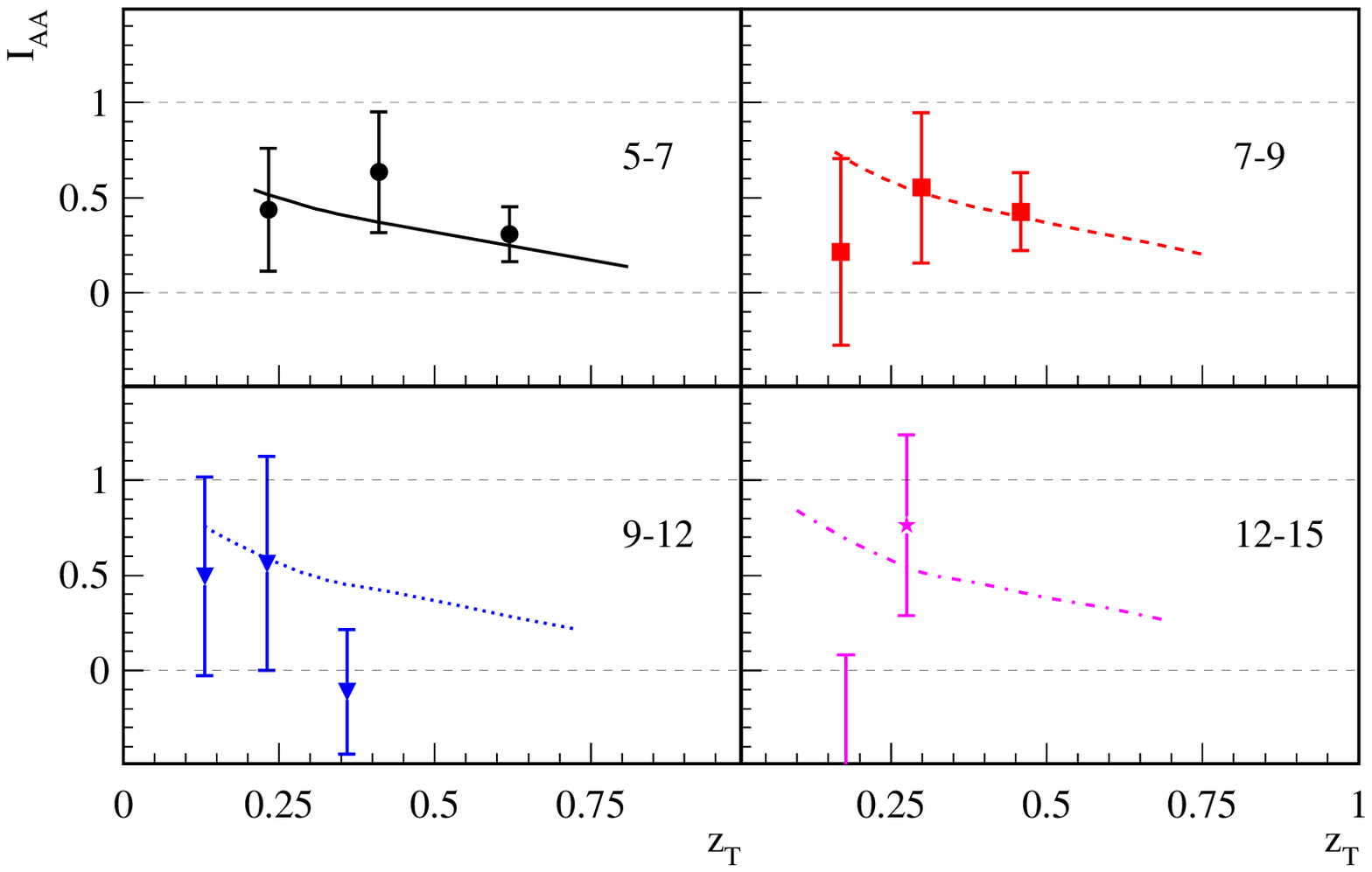, width=4in}
\caption{Nuclear modification factor $I_{AA}(z_T)$ is plotted as a function of $z_T$ at  $\sqrt{S_{NN}}=200$ GeV. 
The numbers on the right top corner in each plots are the range of the photon momentum, while the photon 
and hadron rapidities are integrated over $[-0.35, 0.35]$. The data is from PHENIX~\cite{Adare:2009vd}.}
\label{phenix}
\eef

It is worthwhile to point out that the $\gamma$-triggered light hadron fragmentation function 
$D_{AA}^{\gamma h}(z_T)$ in Au+Au collisions is suppresses relative to that in p+p collisions. This is understandable 
since at LO $D^{\gamma h}(z_T)\propto D_{h/c}(z_T)$ according to Eq.~(\ref{light}), where $D_{h/c}(z)$ is a fast 
falling function of $z$. Since jet quenching can effectively be thought of as a mechanism that probes 
slightly higher $z$ as in  $D_{h/c}(\frac{z}{1-\epsilon})$ according to Eq.~(\ref{quenching}), sampling higher $z$ 
leads to a smaller fragmentation function. 
Thus, the $\gamma$-triggered fragmentation function in A+A collisions will be suppressed compared to 
p+p collisions. We will find this behavior can be altered in the $\gamma$-triggered heavy meson production as we 
will demonstrate below.

In Fig.~\ref{starphenix} (bottom right panel) and Fig.~\ref{phenix}, we compare our calculations of 
$I_{AA}^{\gamma h}$ to the experimental data. Being a ratio of the $\gamma$-triggered fragmentation 
functions in A+A  and p+p collisions, the nuclear modification factor 
$I_{AA}^{\gamma h}(z_T)=D_{AA}^{\gamma h}(z_T)/D_{pp}^{\gamma h}(z_T)$  has much larger error bars. Our results are 
 consistent with the experimental data (within the error bars).

\subsection{Predictions for $\gamma$-triggered light and heavy meson production}
In this subsection we will make predictions for the  $\gamma$-triggered light and heavy meson production at 
both RHIC and LHC energies and will compare the different features of the in-medium modification
depending on the parton mass.  We use $D$ and $B$-meson fragmentation functions as calculated in~\cite{Sharma:2009hn} 
and choose $m=1.3~(4.5)$ GeV for the charm (bottom) quarks. For $\gamma$-triggered heavy meson production 
we have integrated over the relative azimuthal angle $\phi=\phi_h-\phi_\gamma$ in the region: $|\phi-\pi| < \pi/5$ 
in Eq.~(\ref{heavy}), following the experimentally applied cuts~\cite{Adare:2009vd,Abelev:2009gu}. 
\bef
\psfig{file=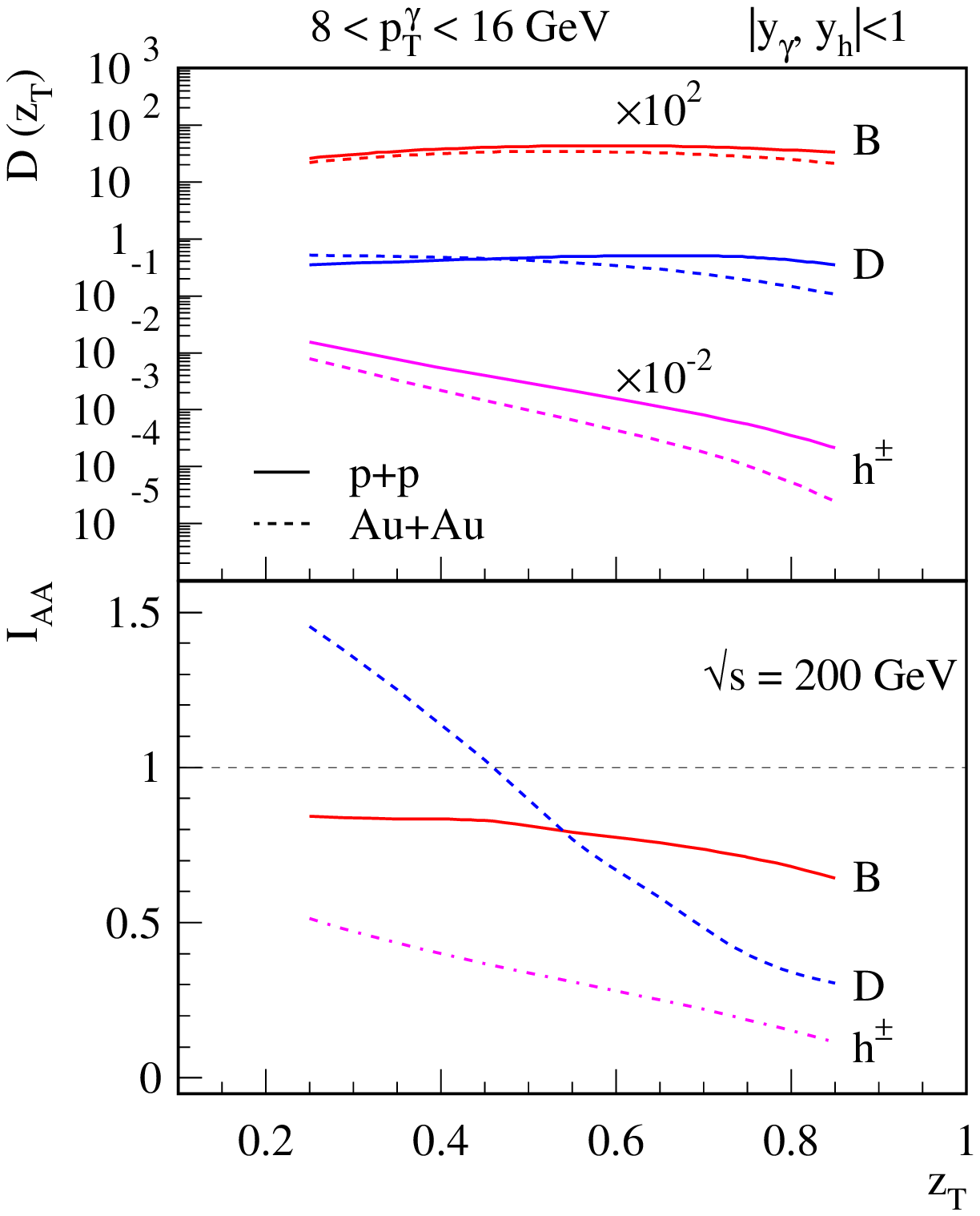, width=2.8in}
\hskip 0.4in
\psfig{file=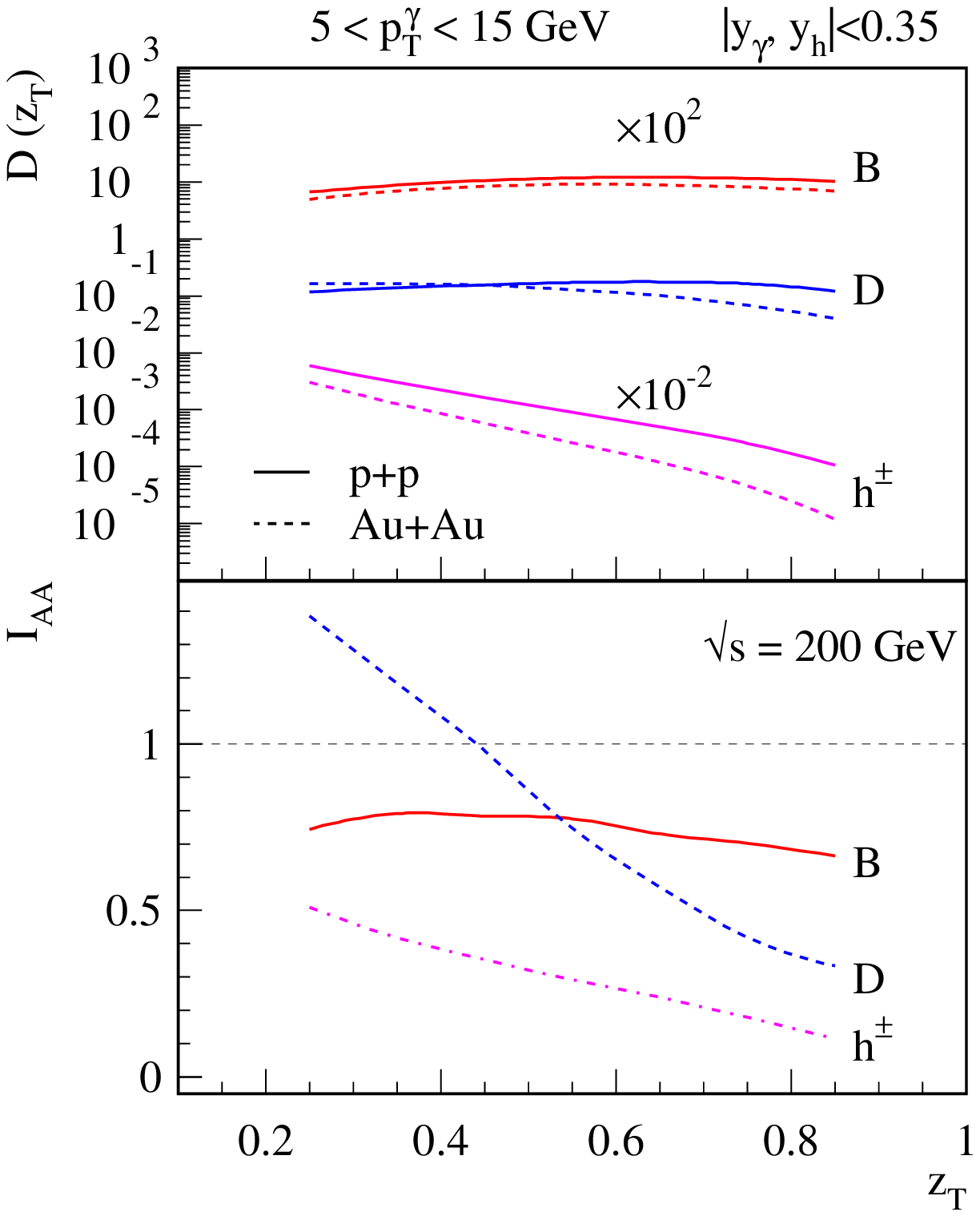, width=2.8in}
\caption{Top panels: predictions for $\gamma$-triggered fragmentation functions $D(z_T)$ where the solid lines 
are for p+p collisions and the dashed lines are for central Au+Au collisions. Bottom panels:  predictions for 
the nuclear modification factor $I_{AA}(z_T)$ where the solid lines are for $B$-meson, the dashed lines are for $D$-meson, 
and the dash-dotted lines are for charged hadrons. The left plot has kinematics similar to STAR, 
while the right plot has  kinematics similar to PHENIX.}
\label{rhic}
\eef

In the upper panels of Fig.~\ref{rhic}  we plot $D(z_T)$ as a function of $z_T$ at $\sqrt{S_{NN}}=200$~GeV 
for both p+p and central Au+Au collisions. Left panels reflect STAR
kinematics and right panels reflect PHENIX kinematics.  As seen clearly in the plots, 
$D(z_T)$ for light hadrons and heavy mesons are very different. For light hadrons, the 
$\gamma$-triggered fragmentation function $D^{\gamma h}(z_T)$ drops very fast as $z_T$ increases, consistent 
with the behavior of light hadron fragmentation function $D_{h/c}(z)$. 
On the other hand, the $\gamma$-triggered heavy meson fragmentation function $D^{\gamma H}(z_T)$ is a relatively 
flat function of $z_T$. Even though in the FFNS scheme $D^{\gamma H}(z_T)$ does not directly correspond to the heavy 
quark to heavy meson decay probability $D_{H/Q}(z)$ itself, it does reflect to some extent the major 
feature of the heavy meson fragmentation function: it grows at lower $z$ and decreases 
at higher $z$. This difference in the fragmentation functions will lead to distinctive difference in the 
nuclear modification factor $I_{AA}$ between light hadron and heavy meson.

\bef
\psfig{file=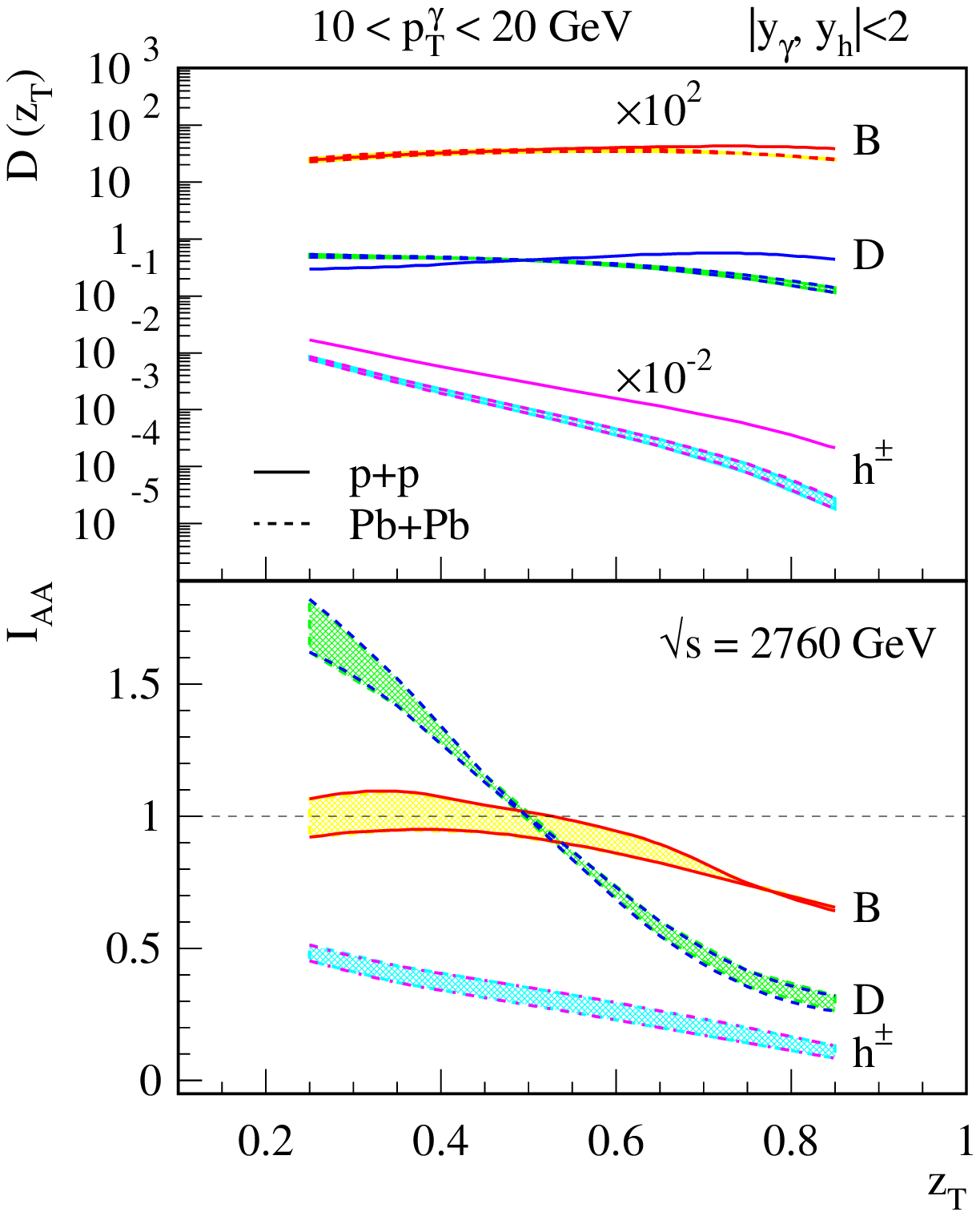, width=2.8in}
\hskip 0.4in
\psfig{file=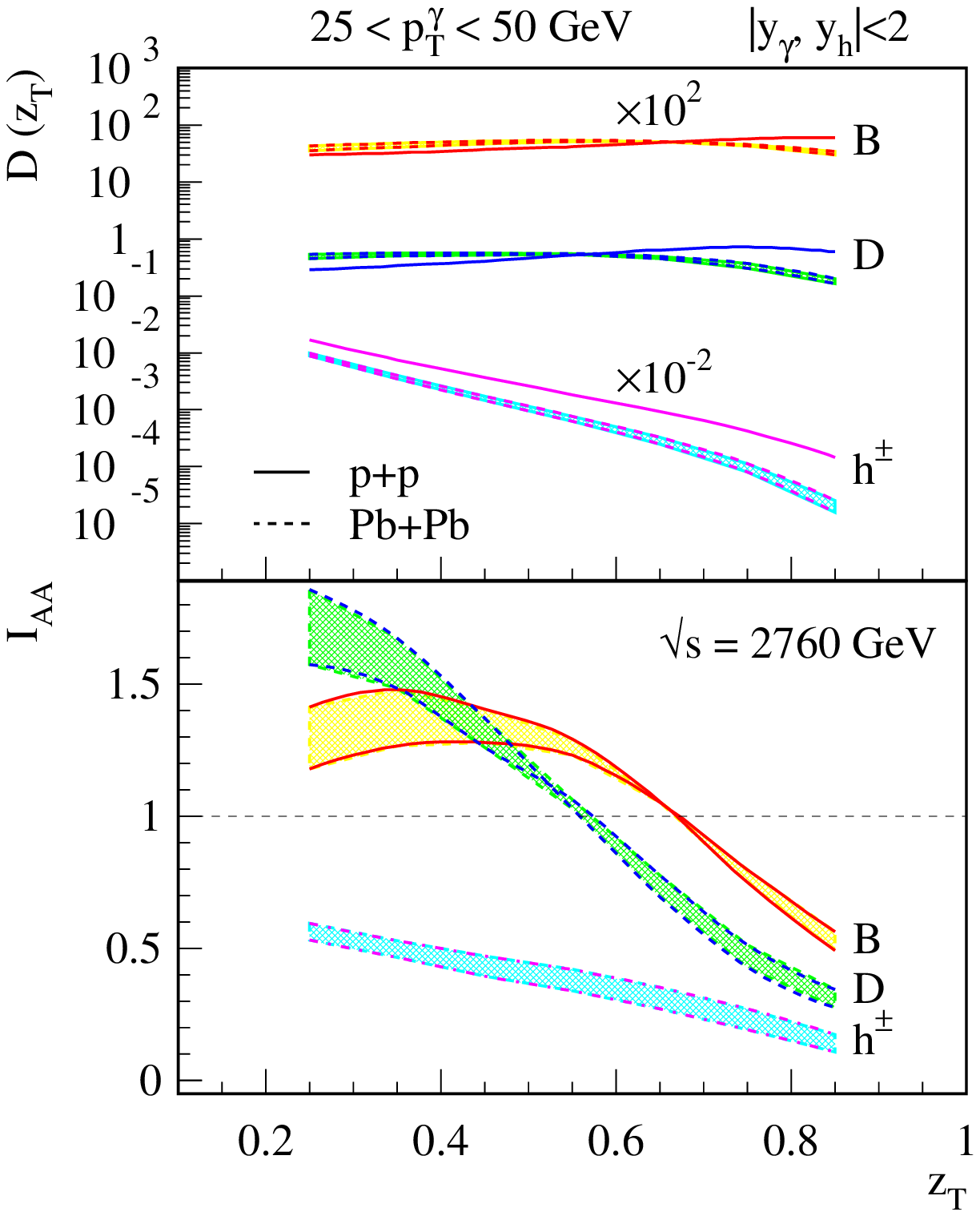, width=2.8in}
\caption{Top panels: predictions for the $\gamma$-triggered fragmentation functions $D(z_T)$, where the solid 
lines are for p+p collisions and the dashed lines are for central Pb+Pb collisions at $\sqrt{S_{NN}}=2.76$~TeV. 
Bottom panels:  predictions for the nuclear modification factor $I_{AA}(z_T)$, where the solid lines are for $B$-meson, 
the dashed lines are for $D$-meson, and the dash-dotted lines are for charged hadrons. We have integrated the 
photon and hadron rapidities over $[-2, 2]$. For the left plot, the photon momentum is integrated over $[10, 20]$~GeV, 
while for the right plot, it has been integrated over $[25, 50]$~GeV.}
\label{lhc}
\eef
In the bottom panels of Fig.~\ref{rhic} we plot  $I_{AA}(z_T)$ as a function of $z_T$ at RHIC for 
$\gamma$-triggered light charged hadrons (dot-dashed), $D$-mesons (dashed) and $B$-mesons (solid), 
respectively. They exhibit very different behavior. For light hadrons, one 
expects $I_{AA}^{\gamma h}<1$ due to jet quenching, as explained above. The magnitude of this suppression 
arrises mainly from the steepness of the light hadron fragmentation function $D_{h/c}(z)$. On the other hand, 
according to Eq.~(\ref{heavy}), for the $\gamma$-triggered heavy meson case $I_{AA}^{\gamma H}$ really depends 
on the whole $z$-integration of the heavy meson fragmentation function $D_{H/Q}(z)$. If the integral is dominated by the 
small-$z$ region, where $D_{H/Q}(z)$ grows with increasing $z$, jet quenching in Eq.~(\ref{quenching}) means 
sampling relatively larger-$z$ and thus larger $D_{H/Q}(z)$. Consequently, one will then have $I_{AA}^{\gamma H}>1$. 
On the other hand, if the integral is dominated by the large-$z$ region, where $D_{H/Q}(z)$ decreases with 
increasing $z$, sampling relatively higher-$z$ means smaller $D_{H/Q}(z)$. One thus has $I_{AA}^{\gamma H}<1$.

One keeps in mind that we have three-particle final state, thus $z_T=p_{T_H}/p_{T_\gamma}$ is not the same 
as the momentum fraction $z$ in heavy meson decay probability $D_{H/Q}(z)$ according to Eq.~(\ref{heavy}). 
Nevertheless, we find that the average $\langle z\rangle$ in the collision does increase as $z_T$ increases. 
Thus, at high $z_T$ where the $z$-integral in $I_{AA}^{\gamma H}$ is dominated by the large $z$ region, one should expect 
that both $B$ and $D$-mesons are suppressed due to jet quenching. The magnitude of the suppression should 
follow $I_{AA}^{\gamma B}>I_{AA}^{\gamma D}>I_{AA}^{\gamma h^\pm}$ if the heavy quark loses less energy than the light 
quark, as predicted by perturbative QCD calculations. On the other hand, in the low $z_T$ region, according to 
our calculation, we find that $I_{AA}^{\gamma D}>1$ for $D$-meson, consistent with our naive expectation, 
that is the low $z$ region dominates the $z$ integral in Eq.~(\ref{heavy}).

However, for $\gamma$-triggered $B$-mesons $I_{AA}^{\gamma B}<1$ for the whole $z_T$ region for the kinematics 
we have chosen at RHIC. This is due to the fact that our $B$-meson fragmentation function is even harder
than the $D$-meson fragmentation function. It drops very fast at high $z$ while increases only slowly at low $z$. 
Thus the nuclear modification from high $z$ (suppression) wins over that from low $z$ (enhancement) for the 
kinematic region we have chosen. Combining the analysis for both the low and the high $z_T$ ends, one immediately 
finds that the nuclear modification $I_{AA}^{\gamma H}$ for $\gamma$-triggered $B$-meson fragmentation function 
is flatter than that for $D$-meson case. Another important  reason for this flatter behavior and the smaller
size of the nuclear modification is the smaller energy loss of $b$-quark in comparison  to $c$-quark. 
Thus, the shape of the nuclear modification for the $\gamma$-triggered fragmentation functions 
can cary valuable information about the properties of medium-induced gluon bremsstrahlung.

In Fig.~\ref{lhc} we give predictions for the $\gamma$-triggered light and heavy meson production in both p+p 
and central Pb+Pb collisions at $\sqrt{S_{NN}}=2.76$~TeV at the LHC. We integrate over both the photon and 
hadron rapidities from -2 to 2. In the left (right) panel, the trigger photon momentum is integrated from 
$10<p_{T_{\gamma}}<20$ GeV ($25<p_{T_{\gamma}}<50$ GeV). The uncertainty band comes from the fact that we have 
included a $\sim 25\%$ uncertainty in the magnitude of the energy loss in our jet quenching calculation. 
The behavior of $I_{AA}$ follows from similar considerations for the relevant kinematics. It is also interesting 
to notice that for low $z_T$  both $I_{AA}^{\gamma B}$ and $I_{AA}^{\gamma D}$ can be larger than unity 
in the chosen kinematic region.

It will be illuminating and important to measure such correlations for both light and heavy meson production 
in p+p and A+A collisions at RHIC and at the LHC. 

\section{Conclusions}
We studied photon-triggered light hadron and heavy meson production in both p+p and A+A collisions. 
We found that the energy loss approach that was successful in describing single hadron production in A+A 
reactions at RHIC could simultaneously describe the STAR and PHENIX experimentally extracted photon-triggered 
light hadron fragmentation functions. 
Using the same theoretical framework, we generalized our formalism to study photon-triggered heavy meson production. 
To take into account the heavy quark mass explicitly, we followed the so-called fixed-flavor-number scheme 
and derived the differential cross section for photon+away-side heavy meson. We found that the nuclear 
modification of photon-tagged heavy meson fragmentation functions in A+A collision is very different 
from that of the photon-tagged light hadron fragmentation functions. This variance was determined to arise  from
the different shape of the decay probabilities for light partons into light hadrons and heavy quarks 
into heavy mesons, respectively. Comparing $D$ and $B$-mesons, we predicted that the nuclear modification 
factor $I_{AA}$ would be flatter for $\gamma$+$B$ production than the one for the $\gamma$+$D$ case. This is 
directly related to the different amount of energy lost by heavy quarks with different mass. Thus, the 
different shape of $I_{AA}$ in $\gamma$+$h$, $\gamma$+$D$, and $\gamma$+$B$ productions can be a sensitive 
and quantitative probe of the strength of medium-induced parton energy loss. Finally, we made detailed 
predictions for both  photon-triggered light and heavy meson at RHIC and at the LHC. We conclude by 
emphasizing once again that a 
comprehensive study of these new final-state channels will provide fresh insight into the details of 
the jet quenching mechanism. 

\section*{Acknowledgments}
We are grateful to Marco Stratmann for useful discussions. This work was supported by the U.S Department of 
Energy under Contract No.~DE-AC02-98CH10886 (Z.K.) and DE-AC52-06NA25396 (I.V.), and in the framework of the 
JET Collaboration. Z.K. thanks the  theoretical division at Los Alamos National Laboratory for its hospitality 
and support where this work was initiated. \\

Note added: we recently became aware of the fact that F. Arleo, I. Schienbein and T. Stavreva are 
working on a similar problem by extending earlier work relevant to p+A collisions~\cite{Stavreva:2010mw}.



\begin{thebibliography}{99}

\bibitem{Gyulassy:1993hr}
  M.~Gyulassy and X.~N.~Wang,
  Nucl.\ Phys.\  B {\bf 420}, 583 (1994)
  [arXiv:nucl-th/9306003].

\bibitem{Baier:1996sk}
  R.~Baier, Y.~L.~Dokshitzer, A.~H.~Mueller, S.~Peigne and D.~Schiff,
  Nucl.\ Phys.\  B {\bf 484}, 265 (1997)
  [arXiv:hep-ph/9608322].
  
\bibitem{Zakharov:1997uu}
  B.~G.~Zakharov,
  JETP Lett.\  {\bf 65}, 615 (1997)
  [arXiv:hep-ph/9704255].
 
\bibitem{Wiedemann:2000za}
  U.~A.~Wiedemann,
  Nucl.\ Phys.\  B {\bf 588}, 303 (2000)
  [arXiv:hep-ph/0005129].

\bibitem{Gyulassy:2000er}
  M.~Gyulassy, P.~Levai and I.~Vitev,
  Nucl.\ Phys.\  B {\bf 594}, 371 (2001)
  [arXiv:nucl-th/0006010].

\bibitem{Wang:2001ifa}
  X.~N.~Wang and X.~f.~Guo,
  Nucl.\ Phys.\  A {\bf 696}, 788 (2001)
  [arXiv:hep-ph/0102230].

\bibitem{Arnold:2002ja}
  P.~B.~Arnold, G.~D.~Moore, L.~G.~Yaffe,
  JHEP {\bf 0206}, 030 (2002).
  [hep-ph/0204343].


\bibitem{Ovanesyan:2011xy}
  G.~Ovanesyan, I.~Vitev,
  [arXiv:1103.1074 [hep-ph]].


\bibitem{d'Enterria:2009am}
  D.~d'Enterria,
  [arXiv:0902.2011 [nucl-ex]].



\bibitem{Adcox:2001jp}
  K.~Adcox {\it et al.}  [PHENIX Collaboration],
  Phys.\ Rev.\ Lett.\  {\bf 88}, 022301 (2002)
  [arXiv:nucl-ex/0109003].

\bibitem{Adler:2002xw}
  C.~Adler {\it et al.}  [STAR Collaboration],
  Phys.\ Rev.\ Lett.\  {\bf 89}, 202301 (2002)
  [arXiv:nucl-ex/0206011].


  
\bibitem{Aamodt:2010jd}
  K.~Aamodt {\it et al.}  [ALICE Collaboration],
  Phys.\ Lett.\  B {\bf 696}, 30 (2011)


\bibitem{Adler:2002tq}
  C.~Adler {\it et al.}  [STAR Collaboration],
  Phys.\ Rev.\ Lett.\  {\bf 90}, 082302 (2003)
  [arXiv:nucl-ex/0210033].

\bibitem{Adare:2010ry}
  A.~Adare {\it et al.}  [The PHENIX Collaboration],
  Phys.\ Rev.\ Lett.\  {\bf 104}, 252301 (2010)
  [arXiv:1002.1077 [nucl-ex]].


\bibitem{Adare:2009vd}
  A.~Adare {\it et al.}  [PHENIX Collaboration],
  Phys.\ Rev.\  C {\bf 80}, 024908 (2009)
  [arXiv:0903.3399 [nucl-ex]].

  
\bibitem{Abelev:2009gu}
  B.~I.~Abelev {\it et al.}  [STAR Collaboration],
  Phys.\ Rev.\  C {\bf 82}, 034909 (2010)
  [arXiv:0912.1871 [nucl-ex]].
  

\bibitem{Salur:2009vz}
  S.~Salur,
  Nucl.\ Phys.\  A {\bf 830}, 139C (2009)
  [arXiv:0907.4536 [nucl-ex]].


\bibitem{Aad:2010bu}
  G.~Aad {\it et al.}  [Atlas Collaboration],
  Phys.\ Rev.\ Lett.\  {\bf 105}, 252303 (2010)
  [arXiv:1011.6182 [hep-ex]].

\bibitem{Chatrchyan:2011sx}
  S.~Chatrchyan {\it et al.}  [CMS Collaboration],
  arXiv:1102.1957 [nucl-ex].


\bibitem{Vitev:2009rd}
  See, for example: 
  H.~Z.~Zhang, Z.~B.~Kang, B.~W.~Zhang and E.~Wang,
  Eur.\ Phys.\ J.\  C {\bf 67}, 445 (2010)
  [arXiv:hep-ph/0609159];
  I.~Vitev and B.~W.~Zhang,
  Phys.\ Rev.\ Lett.\  {\bf 104}, 132001 (2010)
  [arXiv:0910.1090 [hep-ph]];
  Y.~He, I.~Vitev and B.~W.~Zhang,
  arXiv:1105.2566 [hep-ph];
    and the references therein.

\bibitem{Wang:1996yh}
  X.~N.~Wang, Z.~Huang and I.~Sarcevic,
  Phys.\ Rev.\ Lett.\  {\bf 77}, 231 (1996)
  [arXiv:hep-ph/9605213];
  X.~N.~Wang and Z.~Huang,
  Phys.\ Rev.\  C {\bf 55}, 3047 (1997)
  [arXiv:hep-ph/9701227].

\bibitem{Zhang:2009rn}
  H.~Zhang, J.~F.~Owens, E.~Wang and X.~N.~Wang,
  Phys.\ Rev.\ Lett.\  {\bf 103}, 032302 (2009)
  [arXiv:0902.4000 [nucl-th]].

\bibitem{Qin:2009bk}
  G.~Y.~Qin, J.~Ruppert, C.~Gale, S.~Jeon and G.~D.~Moore,
  Phys.\ Rev.\  C {\bf 80}, 054909 (2009)
  [arXiv:0906.3280 [hep-ph]].

\bibitem{Renk:2009ur}
  T.~Renk,
  Phys.\ Rev.\  {\bf C80}, 014901 (2009).
  [arXiv:0904.3806 [hep-ph]].

\bibitem{Dokshitzer:2001zm}
  Y.~L.~Dokshitzer and D.~E.~Kharzeev,
  Phys.\ Lett.\  B {\bf 519}, 199 (2001)
  [arXiv:hep-ph/0106202].

\bibitem{Wicks:2005gt}
  S.~Wicks, W.~Horowitz, M.~Djordjevic and M.~Gyulassy,
  Nucl.\ Phys.\  A {\bf 784}, 426 (2007)
  [arXiv:nucl-th/0512076].

\bibitem{Adil:2006ra}
  A.~Adil and I.~Vitev,
  Phys.\ Lett.\  B {\bf 649}, 139 (2007)
  [arXiv:hep-ph/0611109].

\bibitem{Sharma:2009hn}
  R.~Sharma, I.~Vitev and B.~W.~Zhang,
  Phys.\ Rev.\  C {\bf 80}, 054902 (2009)
  [arXiv:0904.0032 [hep-ph]].

\bibitem{Owens:1986mp}
  J.~F.~Owens,
  Rev.\ Mod.\ Phys.\  {\bf 59}, 465 (1987).
  
\bibitem{Frixione:1994dv}
  S.~Frixione, M.~L.~Mangano, P.~Nason and G.~Ridolfi,
  Phys.\ Lett.\  B {\bf 348}, 633 (1995)
  [arXiv:hep-ph/9412348].

\bibitem{Frixione:1995qc}
  S.~Frixione, P.~Nason and G.~Ridolfi,
  Nucl.\ Phys.\  B {\bf 454}, 3 (1995)
  [arXiv:hep-ph/9506226]; and earlier references given there.

\bibitem{Stratmann:1994bz}
  M.~Stratmann and W.~Vogelsang,
  Phys.\ Rev.\  D {\bf 52}, 1535 (1995).

\bibitem{Berger:1995qe}
  E.~L.~Berger and L.~E.~Gordon,
  Phys.\ Rev.\  D {\bf 54}, 2279 (1996)
  [arXiv:hep-ph/9512343].

\bibitem{Bailey:1996px}
  B.~Bailey, E.~L.~Berger and L.~E.~Gordon,
  Phys.\ Rev.\  D {\bf 54}, 1896 (1996)
  [arXiv:hep-ph/9602373].
  
\bibitem{Stavreva:2009vi}
  T.~P.~Stavreva and J.~F.~Owens,
  Phys.\ Rev.\  D {\bf 79}, 054017 (2009)
  [arXiv:0901.3791 [hep-ph]].
 

\bibitem{Ellis:1987qy}
  R.~K.~Ellis and Z.~Kunszt,
  Nucl.\ Phys.\  B {\bf 303}, 653 (1988).

\bibitem{Vitev:2002pf}
  See, for example: I.~Vitev and M.~Gyulassy,
  Phys.\ Rev.\ Lett.\  {\bf 89}, 252301 (2002)
  [arXiv:hep-ph/0209161];
  X.~N.~Wang,
  Phys.\ Rev.\  C {\bf 70}, 031901 (2004)
  [arXiv:nucl-th/0405029];
  A.~Adil and M.~Gyulassy,
  Phys.\ Lett.\  B {\bf 602}, 52 (2004)
  [arXiv:nucl-th/0405036].

\bibitem{Gyulassy:2000fs}
  M.~Gyulassy, P.~Levai and I.~Vitev,
  Phys.\ Rev.\ Lett.\  {\bf 85}, 5535 (2000)
  [arXiv:nucl-th/0005032].


\bibitem{Vitev:2007ve}
  I.~Vitev,
  Phys.\ Rev.\  C {\bf 75}, 064906 (2007)
  [arXiv:hep-ph/0703002].


\bibitem{Neufeld:2010fj}
  R.~B.~Neufeld, I.~Vitev and B.~W.~Zhang,
  Phys.\ Rev.\  C {\bf 83}, 034902 (2011)
  [arXiv:1006.2389 [hep-ph]].

\bibitem{Pumplin:2002vw}
  J.~Pumplin, D.~R.~Stump, J.~Huston, H.~L.~Lai, P.~M.~Nadolsky and W.~K.~Tung,
  JHEP {\bf 0207}, 012 (2002)
  [arXiv:hep-ph/0201195].

\bibitem{deFlorian:2007aj}
  D.~de Florian, R.~Sassot and M.~Stratmann,
  Phys.\ Rev.\  D {\bf 75}, 114010 (2007)
  [arXiv:hep-ph/0703242].


\bibitem{Stavreva:2010mw}
  T.~Stavreva, I.~Schienbein, F.~Arleo, K.~Kovarik, F.~Olness, J.~Y.~Yu, J.~F.~Owens,
  JHEP {\bf 1101}, 152 (2011).
  [arXiv:1012.1178 [hep-ph]].


                                  
\end{thebibliography}
\end{document}